\documentclass[reprint,3p]{elsarticle} 

\usepackage{amsmath}
\usepackage{lineno,hyperref}

\journal{Journal of the Mechanics and Physics of Solids}

\biboptions{authoryear}

\usepackage{bm}
\usepackage{dsfont}

\begin{document}

\begin{frontmatter}

\title{Strain localization regularization and patterns formation in
rate-dependent plastic materials with multiphysics coupling}

\author[1]{Antoine B. Jacquey\corref{corrauthor}}
\ead{ajacquey@mit.edu}

\author[2]{Hadrien Rattez}
\author[3]{Manolis Veveakis}

\address[1]{Department of Civil and Environmental Engineering, Massachusetts Institute of Technology, MA, USA.}
\address[2]{Civil and Environmental Engineering, UCLouvain, Belgium}
\address[3]{Civil and Environmental Engineering Department, Duke University, NC, USA.}
\cortext[corrauthor]{Corresponding author}

\begin{abstract}
  Strain localization is an instability phenomenon occurring in deformable solid
  materials which undergo dissipative deformation mechanisms. Such instability
  is characterized by the localization of the displacement or velocity fields in
  a zone of finite thickness and is generally associated with the failure of
  materials. In several fields of material engineering and natural sciences,
  estimating the thickness of localized deformation is required to make accurate
  predictions of the evolution of the physical properties within localized
  strain regions and of the material strength. In this context, scientists and
  engineers often rely on numerical modeling techniques to study strain
  localization in solid materials. However, classical continuum theory for
  elasto-plastic materials fails at estimating strain localization thicknesses
  due to the lack of an internal length in the model constitutive laws. In this
  study, we investigate at which conditions multiphysics coupling enables to
  regularize the problem of strain localization using rate-dependent plasticity.
  We show that coupling the constitutive laws for deformation to a single
  generic diffusion-reaction equation representing a dissipative state variable
  can be sufficient to regularize the ill-posed problem under some conditions on
  the softening parameters in the plastic potential. We demonstrate in these
  cases how rate-dependent plasticity and multiphysics coupling can lead to
  material instabilities depicting one or several internal length scales
  controlled by the physical parameters resulting in the formation of regular or
  erratic patterns. As we consider a general form of the equations, the results
  presented in this study can be applied to a large panel of examples in the
  material engineering and geosciences communities.
\end{abstract}


\end{frontmatter}


\section{Introduction}

Solid materials can undergo irreversible deformation mechanisms when subject to
large deformation and consequently to large stresses. Such irreversible
deformation is usually associated with the dissipation of some energy quantity
which may depend on a set of coupled state variables such as temperature and
pressure. In a large variety of solid materials undergoing irreversible
deformation, the displacement and velocity fields can localize in finite
thicknesses which are commonly described as strain localization phenomena.
Understanding the conditions which lead to localization instabilities and how
they evolve has become of relevance in several fields of material engineering
and natural sciences as localization instabilities can drastically alter the
physical properties of the solid materials and ultimately lead to its failure.
Observations of localization instabilities are well documented by laboratory
studies for a large variety of materials, including: (i) metals and alloys
\citep{Tresca1878, Antolovich2014}; (ii) biomaterials
\citep{Richman1975,Budday2014,Molnar2018}; or (iii) geomaterials
\citep{Desrues1996, Baud2004a, Fossen2007a}. To explain these observations, the
scientific community has put forward several theoretical frameworks in the last
decades \citep{Rudnicki1975a,Issen2000}. One major outcome of this theoretical
effort was to recognize that Cauchy-Boltzmann mechanical formulation, which is
usually adopted to describe the deformation of solids fails at describing strain
localization phenomena as it predicts localized thicknesses of zero-size
\citep{Iordache1998,Rattez2018}. Such formulations are therefore said to be
ill-posed as they induce mesh-dependency in finite-element analyses accompanied
with convergence defects \citep{deBorst1993,Vardoulakis1995,Rattez2018a}. To
mitigate this effect, several regularization techniques are found in the
literature, such as higher-order continuum theories
\citep{Muhlhaus1987,Pijaudier1987,Vardoulakis1991,Forest2005}, multiphysics
coupling \citep{Benallal2004,Benallal2005,Rice2014,Veveakis2014c} or viscous
regularization \citep{Needleman1988,Loret1990,Prevost1990,Wang1996}.

The concept of viscous regularization of the strain localization problem was
first introduced by \citet{Needleman1988} and later followed by
\citet{Prevost1990,Loret1990} who described by means of numerical tests how
rate-dependency resolves the mesh-dependency of strain localization for
softening materials under quasi-static and dynamic loading conditions. As
explained in \citep{Bazant2002}, the rate dependency introduces a characteristic
length scale -- only when the inertial terms are considered -- which allows to
prevent the change of type of partial differential equations that occurs with
strain softening, rate independent materials and to regularize the strain
localization problem. However, this regularization presents peculiarities
compared to other regularization methods like high order continuum theories. Its
effect tends to disappear in time if the time scale of the problem is larger
than the relaxation time associated with the viscous terms. Thus, this viscous
regularization can only be used within a narrow range of time delays and rates
\citep{Bazant2002}. \citet{Benallal2008} later investigated the conditions in
terms of loading conditions, viscosity and numerical time discretization for the
regularization by viscous terms to be effective. Moreover, this rate-dependent
regularization is more sensitive to the imperfections triggering the
localization process than other methods \citep{Molinari1987,Belytschko1991}.
\citet{Wang1996} have shown that the size of the strain localization zone is
controlled by the minimum of the imperfection length scale and the viscous
length scale, that is why recent numerical study have introduced a large
perturbation length scale in their numerical setting to study localization in
geomaterials \citep{Duretz2014,Duretz2019,Jacquey2020}. In his seminal paper,
\citet{Needleman1988} also noticed that the length scale of strain localization
is related to heterogeneities or imperfections and speculated that considering
additional coupling to diffusive processes would integrate a physical length
scale into the constitutive laws of deformation. This coupling of
rate-dependency and a diffusion equation has been applied to study the
phenomenon of adiabatic shear bands \citep{Batra1991,McAuliffe2013} for the
particular case of the temperature-dependent mechanical behavior and it has been
shown that for a variety of imperfections the shear-band structure depends
almost entirely on the imposed nominal strain rate and material parameters
\citep{Bayliss1994}. However, all these studies have shown numerically the
regularization by the viscous-diffusion coupling for a single or limited set of
parameters, but it remains unclear under which conditions the coupling of a
diffusion equation regularizes the problem of strain localization or how the
viscous and diffusive characteristic lengths interact to control the
localization for problems involving a small relaxation time compared to the
viscous time scale.

To address these questions, we consider in this study a general combination of
multiphysics coupling and viscous terms to investigate the regularization of the
ill-posed problem. We performed a linear stability analysis of the resulting
system of equations and demonstrate the conditions necessary to regularize the
strain localization problem. We have identified several strain localization
regimes based on the dependency of the plastic yield to accumulated plastic
strain and coupled dissipative variable. Two specific regimes are covered in
this study: one for which strain localization regularization induces the
formation of regular patterns controlled by standing waves and another where the
regularization can induce complex patterns controlled by the interplay of
standing and propagating waves. We also present the dynamic evolution of such
systems by means of finite-element simulations.

\section{Problem formulation}
\label{sec:gov_eq}

In the following, we describe the mathematical formulation of the
one-dimensional deformation of a solid material using a rate-dependent
plasticity model coupled to a diffusion-reaction equation. Materials subject to
pure shear, pure compression or uniaxial deformation can be reduced to
one-dimensional problems but more complex loading conditions would require to
account for additional dimensions. We make use of a classical additive splitting
of the total strain (\(\gamma\)) into an elastic (\(\gamma^{el}\)) and a plastic
component (\(\gamma^{p}\)):

\begin{equation}
  \gamma = \frac{\partial u}{\partial x} = \gamma^{el} + \gamma^{p}
\end{equation}

The governing equation for the displacement variable \(u\) in one
dimension is obtained based on the balance of momentum:

\begin{equation}
  \rho \frac{\partial^{2} u}{\partial t^{2}} - \frac{\partial \tau}{\partial x}
  = \rho \frac{\partial^{2} u}{\partial t^{2}} - G \frac{\partial^{2}
  u}{\partial x^{2}}  + G \frac{\partial \gamma^{p}}{\partial x} = 0
\label{eq:motion}
\end{equation}

\noindent where \(G\) is the elastic modulus and \(\rho\) the material density.
Equation~\ref{eq:motion} is similar to a wave equation with an additional term
depending on the plastic behavior of the material. Equation~\ref{eq:motion} can
also be expressed by considering the velocity \(v = \frac{\partial u}{\partial
t}\) as the primary variable for deformation:

\begin{equation}
  \frac{1}{c^{2}} \frac{\partial^{2} v}{\partial t^{2}} - \frac{\partial^{2}
  v}{\partial x^{2}} + \frac{\partial \dot{\gamma}^{p}}{\partial x} = 0,
\label{eq:motion_vel}
\end{equation}

\noindent where \(c = \sqrt{\frac{G}{\rho}}\) is the elastic wave velocity and
\(\dot{\gamma}^{p}\) the plastic strain rate. The plastic strain is generally
considered as an internal variable and not a state variable. Additionally, we
consider a coupled variable \(\theta\) (here considered dimensionless for sake
of generality) governed by a diffusion-reaction type of equation:

\begin{equation}
  \frac{\partial \theta}{\partial t} - \kappa \frac{\partial^{2} \theta}{\partial
  x^{2}} - \alpha \dot{\gamma}^{p} = 0,
\label{eq:diff}
\end{equation}

\noindent where \(\kappa\) is the diffusivity of the material and \(\alpha\) a
coefficient controlling the release of mechanical energy. We keep the coupled
variable \(\theta\) as a generic variable in this study as it can represent a
large variety of processes depending of the type of material considered, the
spatial and temporal scales at which we consider the deformation of the material
and the type of forcing conditions the material is subject to. In a general
sense, Equation~\ref{eq:diff} represents the diffusion of an internal quantity
which is relevant for the deformation of the material. The variable \(\theta\)
could therefore represent physical quantities such as temperature of the
material; the saturation or the fluid pressure in the case of fully-saturated
porous media; the mass concentration of a specific mineral; or the
microstructure organization representing a phase transition. This variable can
also be interpreted as a non-local version of the accumulated plastic strain, in
the form of a damage or breakage variable often adopted for granular media
\citep{Kamrin2019}. The source term \(\alpha \dot{\gamma}^{p}\) is chosen to be
generic, representing effects like the dissipation rate in the case of
\(\theta\) being the temperature, the volumetric plastic rate of deformation in
mass balance considerations, the deviatoric plastic strain rate in the case of
\(\theta\) being a generalized damage variable expressing shear failure, to
mention only a few examples. Here for sake of simplicity, we consider \(\alpha =
1\) but in a general sense this coefficient can be a function of stress, elastic
strain or other state variables depending on the physics considered.

We consider the onset of plastic deformation as being controlled by the
following yield function formulated in terms of stress:

\begin{equation}
  \mathcal{F} = \tau - \tau_{y}\left(\gamma^{p}, \theta\right)
\label{eq:yield}
\end{equation}

\noindent where \(\tau\) is the stress and \(\tau_{y}\) the yield stress
expressed as a function of the two variables \(\gamma^{p}\) and \(\theta\).
Assuming a rate-dependent plastic model or viscoplastic (overstress) rheology,
the flow law for the plastic strain is expressed as:

\begin{equation}
  \dot{\gamma}^{p} = \frac{\langle \mathcal{F}
  \rangle}{\eta} \frac{\partial \mathcal{F}}{\partial \tau},
\end{equation}

\noindent where \(\eta\) is the material viscosity and \(\langle x \rangle =
\frac{x + \lvert x \rvert}{2}\) denotes the Macauley brackets. Using the
expression of the function introduced in
Equation~\ref{eq:yield}, one can obtain the following flow rule
for the viscoplastic strain:

\begin{equation}
  \dot{\gamma}^{p} = \frac{G}{\eta} \langle  \bar{\tau} -
  \bar{\tau}_{y}\left(\gamma^{p}, \theta\right)\rangle,
\label{eq:plastic_strain}
\end{equation}

\noindent where \(\bar{\tau} = \frac{\tau}{G}\) and \(\bar{\tau}_{y} =
\frac{\tau_{y}}{G}\) are the dimensionless stress and yield stress. Assuming a
material already undergoing plastic deformation, we can assume the stress is
larger than the yield stress and therefore simplify
Equation~\ref{eq:plastic_strain} by removing the Macauley brackets. The plastic
strain rate can be eliminated from the governing equations by combining
Equations~\ref{eq:plastic_strain} and~\ref{eq:motion_vel} and considering
\(\Theta = \frac{\partial \theta}{\partial x}\) as the primary coupled variable
and combining Equations~\ref{eq:plastic_strain} and~\ref{eq:diff} which results
in:

\begin{eqnarray}
  \frac{1}{c^{2}} \frac{\partial^{2} v}{\partial t^{2}} + \frac{G}{\eta c^{2}}
  \frac{\partial v}{\partial t} - \frac{\partial^{2} v}{\partial x^{2}} -
  \frac{G}{\eta} \frac{\partial \bar{\tau}_{y}}{\partial x} &=& 0, \nonumber \\
  \frac{\partial \Theta}{\partial t} - \kappa \frac{\partial^{2}
  \Theta}{\partial x^{2}} - \frac{G}{\eta c^{2}} \frac{\partial v}{\partial t} +
  \frac{G}{\eta} \frac{\partial \bar{\tau}_{y}}{\partial x} &=& 0.
\label{eq:final}
\end{eqnarray}

We introduce
the following dimensionless spatial and temporal variables \(\bar{x}\) and
\(\bar{t}\) respectively depending on the material properties and loading
conditions:

\begin{equation}
  \bar{t} = \dot{\gamma}_{0} t, \quad \bar{x} = \frac{\dot{\gamma}_{0}}{c} x,
  \quad \bar{u} = \frac{\dot{\gamma}_{0}}{c} u, \quad \bar{v} = \frac{v}{c},
  \quad \bar{\Theta} = \frac{c}{\dot{\gamma}_{0}} \Theta
\label{eq:dimensionless}
\end{equation}

\noindent where \(\dot{\gamma}_{0}\) is the loading or background strain rate.
Making use of the dimensionless quantities introduced in
Equation~\ref{eq:dimensionless}, the final set of governing equations can be
simplified as:

\begin{eqnarray}
  \frac{\partial^{2} \bar{v}}{\partial \bar{t}^{2}} + \xi \frac{\partial
  \bar{v}}{\partial \bar{t}} - \frac{\partial^{2} \bar{v}}{\partial \bar{x}^{2}}
  - \xi \frac{\partial \bar{\tau}_{y}}{\partial \bar{x}} &=& 0, \nonumber \\
  \frac{\partial \bar{\Theta}}{\partial \bar{t}} - \frac{\chi}{\xi}
  \frac{\partial^{2} \bar{\Theta}}{\partial \bar{x}^{2}} - \xi \frac{\partial
  \bar{v}}{\partial \bar{t}} + \xi \frac{\partial \bar{\tau}_{y}}{\partial
  \bar{x}} &=& 0.
\label{eq:final_dimensionless}
\end{eqnarray}

\noindent where we introduced the following dimensionless parameters:

\begin{equation}
  \chi = \frac{\kappa \rho}{\eta}, \quad
  \xi = \frac{G}{\dot{\gamma}_{0} \eta}.
\end{equation}

The parameter \(\chi\) represents the ratio of physical (or coupling)
diffusivity to momentum diffusivity (or kinematic viscosity). Its value is
governed by the physical properties of the material. The parameter \(\xi\) is
the ratio of the loading time scale to the viscous time scale. This parameters
depends both on the boundary condition and the material physical properties. The
viscosity of the material \(\eta\) influences both of these dimensionless
parameters. However, the effective diffusivity of the coupled variable
\(\bar{\Theta}\) expressed as the ratio of the two introduced dimensionless
parameters is independent of the viscosity.

\section{Linear stability Analysis}
\label{sec:lsa}

\subsection{Governing equations for the perturbations}

To analyze the stability of deformation mechanisms, it is common to consider the
loss of ellipticity of the acoustic tensor as a condition for strain
localization. However, when deformation is coupled to a set of physical
variables, it is necessary to consider the evolution of these coupled quantities
and take into account how their changes impact deformation. For that purpose, it
is more convenient to perform a linear stability analysis of the resulting
system of coupled equations. In the absence of multiphysics coupling, the
results from a linear stability analysis yield the same stability condition as
the one obtained by considering the loss of ellipticity of the acoustic tensor
\citep{Rattez2018}.

In the following, we investigate the onset of strain localization in the system
described by Equations~\ref{eq:final_dimensionless} by means of a linear
stability analysis. We consider the set of solutions described by a small
perturbation (noted with superscript \(\star\)) around a steady-state solution
(noted with subscript \(0\)):

\begin{eqnarray}
  \bar{u} \left(\bar{x}, \bar{t}\right) &=& \bar{u}_{0} \left(\bar{x}\right) +
  \bar{u}^{\star}\left(\bar{x}, \bar{t}\right), \nonumber \\
  \bar{\Theta} \left(\bar{x}, \bar{t}\right) &=& \bar{\Theta}_{0}
  \left(\bar{x}\right) + \bar{\Theta}^{\star} \left(\bar{x}, \bar{t}\right).
\end{eqnarray}

For sake of simplicity, we also make use of the following notation for the
velocity perturbation \(\bar{v}^{\star} = \frac{\partial
\bar{u}^{\star}}{\partial \bar{t}}\). The perturbation solutions satisfy the
following linearized version of Equation~\ref{eq:final_dimensionless} given by:

\begin{eqnarray}
  \frac{\partial^{2} \bar{v}^{\star}}{\partial \bar{t}^{2}} + \xi \left(1 +
  h_{\gamma}\right) \frac{\partial \bar{v}^{\star}}{\partial \bar{t}} -
  \frac{\partial^{2} \bar{v}^{\star}}{\partial \bar{x}^{2}} - \xi h_{\gamma}
  \frac{\partial^{2} \bar{u}^{\star}}{\partial \bar{x}^{2}} - \xi h_{\theta}
  \bar{\Theta}^{\star} &=& 0, \nonumber \\
  \frac{\partial \bar{\Theta}^{\star}}{\partial \bar{t}} - \frac{\chi}{\xi}
  \frac{\partial^{2} \bar{\Theta}^{\star}}{\partial \bar{x}^{2}} - \xi \left(1 +
  h_{\gamma}\right) \frac{\partial \bar{v}^{\star}}{\partial \bar{t}} + \xi
  h_{\gamma} \frac{\partial^{2} \bar{u}^{\star}}{\partial \bar{x}^{2}} + \xi
  h_{\theta} \bar{\Theta}^{\star} &=& 0,
\label{eq:final_linearized}
\end{eqnarray}

\noindent where we made use of a Taylor expansion for the expression of the
yield stress derivative: \(\frac{\partial \bar{\tau}_{y}}{\partial \bar{x}}
\left(\gamma^{p}, \theta\right) = \frac{\partial \bar{\tau}_{y}}{\partial
\bar{x}} \left(\gamma^{p}_{0}, \theta_{0}\right) + h_{\gamma} \frac{\partial
\gamma^{p\star}}{\partial \bar{x}} + h_{\theta} \bar{\Theta}^{\star}\) where
\(h_{\gamma} = \frac{\partial \bar{\tau}_{y}}{\partial \gamma^{p}}
\left(\gamma^{p}_{0}, \theta_{0}\right)\) and \(h_{\theta} = \frac{\partial
\bar{\tau}_{y}}{\partial \theta} \left(\gamma^{p}_{0}, \theta_{0}\right)\) are
the effective moduli defined as the derivative of the yield stress with respect
to the plastic strain and coupled variable estimated at the steady-state
conditions. As we do not consider here a specific yield stress expression, we
will consider different cases depending on the sign (hardening/softening) and
the absolute values of these two plastic moduli. The results of the following
linear stability analysis are therefore relevant for any non-linear system
(independently from the expression of the yield stress) which can be linearized
in the form of Equation~\ref{eq:final_linearized}. We do not investigate the
expressions of the steady-state solutions in this study as we do not consider a
particular yield stress expression. The dependency of the steady-state solutions
are rather captured by exploring the different values of the plastic moduli.
In the following, we therefore assume the existence of a steady-state solution
and analyze the evolution of the perturbations.

Applying a space Fourier transform to the set of
Equations~\ref{eq:final_linearized}, it can be demonstrated that the
perturbations can be expressed as:

\begin{eqnarray}
  \bar{u}^{\star}\left(\bar{x}, \bar{t}\right) &=& \bar{u}^{\star} \exp\left(i k
  \bar{x} + s \bar{t} \right), \nonumber \\
  \bar{\Theta}^{\star}\left(\bar{x}, \bar{t}\right) &=& \bar{\Theta}^{\star}
  \exp\left(i k \bar{x} + s \bar{t} \right),
\label{eq:perturbation}
\end{eqnarray}

\noindent where \(\bar{u}^{\star}\) and \(\bar{\Theta}^{\star}\) on the
right-end side of Equation~\ref{eq:perturbation} are independent of the
dimensionless space and time variables. The parameters \(k\) and \(s\) are the
wave number and the rate of growth of the perturbation (or Lyapunov exponent)
respectively. By substituting the expressions of the solutions
(Equations~\ref{eq:perturbation}) into the governing
Equations~\ref{eq:final_linearized}, one can obtain the following matrix system
governing the perturbations amplitudes:

\begin{equation}
  \begin{bmatrix}
    s^{2} \left(s + \xi \left(1 + h_{\gamma}\right)\right) + k^{2} \left(s + \xi
    h_{\gamma}\right) & - \xi h_{\theta} \\
    -\xi \left(1 + h_{\gamma}\right) s^{2} - \xi h_{\gamma} k^{2} & s +
    -\frac{\chi}{\xi} k^{2} + \xi h_{\theta}
  \end{bmatrix}
  \begin{bmatrix}
    \bar{u}^{\star} \\
    \bar{\Theta}^{\star}
  \end{bmatrix}
  =
  \begin{bmatrix}
    0 \\
    0
  \end{bmatrix}
\label{eq:lsa_cond}
\end{equation}

In this study, we consider the condition for strain localization when the above
system of equations becomes unstable in the Lyapunov sense (when \(s>0\)).
The strain localization condition in the context of the coupled system described
by Equation~\ref{eq:lsa_cond} reads:

\begin{equation}
  \det \begin{bmatrix}
    s^{2} \left(s + \xi \left(1 + h_{\gamma}\right)\right) + k^{2} \left(s + \xi
    h_{\gamma}\right) & - \xi h_{\theta} \\
    -\xi \left(1 + h_{\gamma}\right) s^{2} - \xi h_{\gamma} k^{2} & s +
    -\frac{\chi}{\xi} k^{2} + \xi h_{\theta}
  \end{bmatrix} = 0,
\label{eq:strain_loc_cond}
\end{equation}

\noindent which gives some solutions depending on the dimensionless rate of
growth of the perturbation \(s\) and on \(k = \frac{2 \pi}{\lambda}\) where
\(\lambda\) is the dimensionless wavelength of the perturbation and on the
different physical parameters of the system (\(h_{\pi}\), \(h_{\theta}\),
\(\chi\) and \(\xi\)).

The characteristic polynomial for strain localization is a fourth order
expression in \(s\), whose solutions depend on the dimensionless wavelength and
the physical properties. We analyze the stability of the system in terms of the
solution of \(s\). From the expression of the perturbations
(Equations~\ref{eq:perturbation}), it is clear that any perturbations will
vanish over time if the rate of growth of the perturbation \(s\) is negative,
characterizing a stable regime. On the other hand, if the rate of growth is
positive, the perturbations will increase exponentially over time and therefore
refers to an unstable regime. Solutions of Equation~\ref{eq:strain_loc_cond} can
also lead to complex expressions of the rate of growth \(s\). In that case,
using the expression of the perturbations, it is also clear that the stability
of the system, thus the exponential increase of decay of the perturbation
amplitudes depend only on the sign of the real part of the rate of growth of the
perturbation as illustrated by the following equation:

\begin{equation}
  X^{\star} \left(\bar{x}, \bar{t}\right) = X^{\star} \exp\left[ik\bar{x} +
  \left(Re\left(s\right) + i Im\left(s\right) \right)\bar{t}\right] = X^{\star}
  \exp\left(Re\left(s\right) \bar{t}\right) \exp\left[i\left(k \bar{x} +
  Im\left(s\right)\bar{t}\right)\right],
\end{equation}

\noindent where \(Re\left(\cdot\right)\) and \(Im\left(\cdot\right)\) are the
real and imaginary parts of a complex number and \(X^{\star}\) is the
perturbation amplitudes of the main variables.

\subsection{Effect of multiphysics coupling for rate-dependent materials}

\begin{figure}
  \centering
  \includegraphics[width=0.5\textwidth]{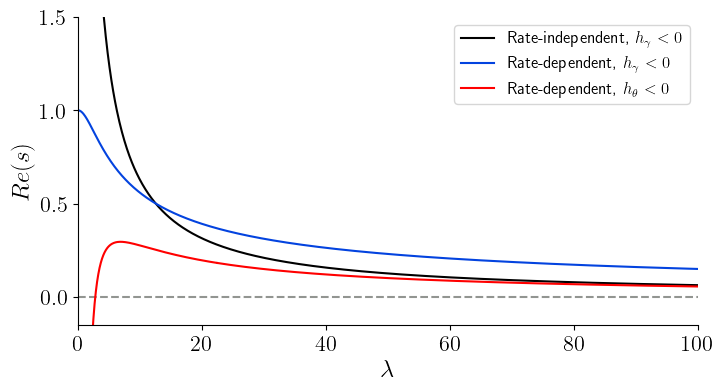}
  \caption{\label{fig:uncoupled}Results of the linear stability analysis for a
  rate-independent plastic model subject to internal softening (black line), for
  a rate-dependent plastic model subject to internal softening (blue line) and
  for a rate-dependent plastic model subject to coupling softening.}
\end{figure}

In this section, we will investigate the role of multiphysics coupling as a mean
to regularize the strain localization problem. To put in perspective such role
of the multiphysics coupling, we illustrate in Figure~\ref{fig:uncoupled} the
results obtained from linear stability analysis which emphasize the role of
rate-dependency and multiphysics coupling. The solid black line corresponding to
a rate-independent material subject to internal softening and no multiphysics
coupling illustrate the ill-posedness of the strain localization problem for
Cauchy continua. The rate of growth of the perturbation tends to infinity for
zero wavelength characterizing a strain localization which is neither
regularized in time nor in space. This leads to convergence problems for
numerical simulators and mesh-sensitive strain localization results. As the
dominant wavelength for strain localization is zero, deformation localizes in
bands of the size of one element which prevents any reliable estimation of the
localization thickness. The blue line, corresponding to a similar material but
including rate-dependency (obtained by setting \(h_{\gamma} < 0\) and
\(h_{\theta} = 0\) in Equation~\ref{eq:final_linearized}) illustrates the
partial regularization of rate-dependent rheologies. The rate of growth of the
perturbation reaches a bounded maximum for a zero wavelength. The viscous terms
introduced by the rate-dependent constitutive laws allow to regularize the
system in time (rate of growth does not tend towards infinity) but fail to
regularize the strain localization problem in space as the regularization by
rate dependence observed numerically disappears over time. It is thus only
effective if used for loading timescales shorter than material relaxation time
and the shear band size depends strongly on the imperfection wavelength imposed
to trigger the localization. These findings have been covered in some details by
\citet{Bazant2002,Stefanou2019}. The red solid line on the other hand
corresponds to a rate-dependent material only subject to multiphysics coupling
(\(h_{\theta} < 0\)) without internal plastic hardening/softening (\(h_{\gamma}
= 0\)). Such setting allows to isolate the impact of multiphysics coupling on
the strain localization regularization. The rate of growth of the perturbation
reaches a maximum for a positive and bounded wavelength. The value of this
wavelength (noted \(\hat{\lambda}\)) and corresponding to the maximum of the
rate of growth of the perturbation allows to estimate the spatial extend of
localized displacement or velocity fields or the spacing between regular
patterns produced by strain localization.

\begin{figure}
  \centering
  \includegraphics[width=\textwidth]{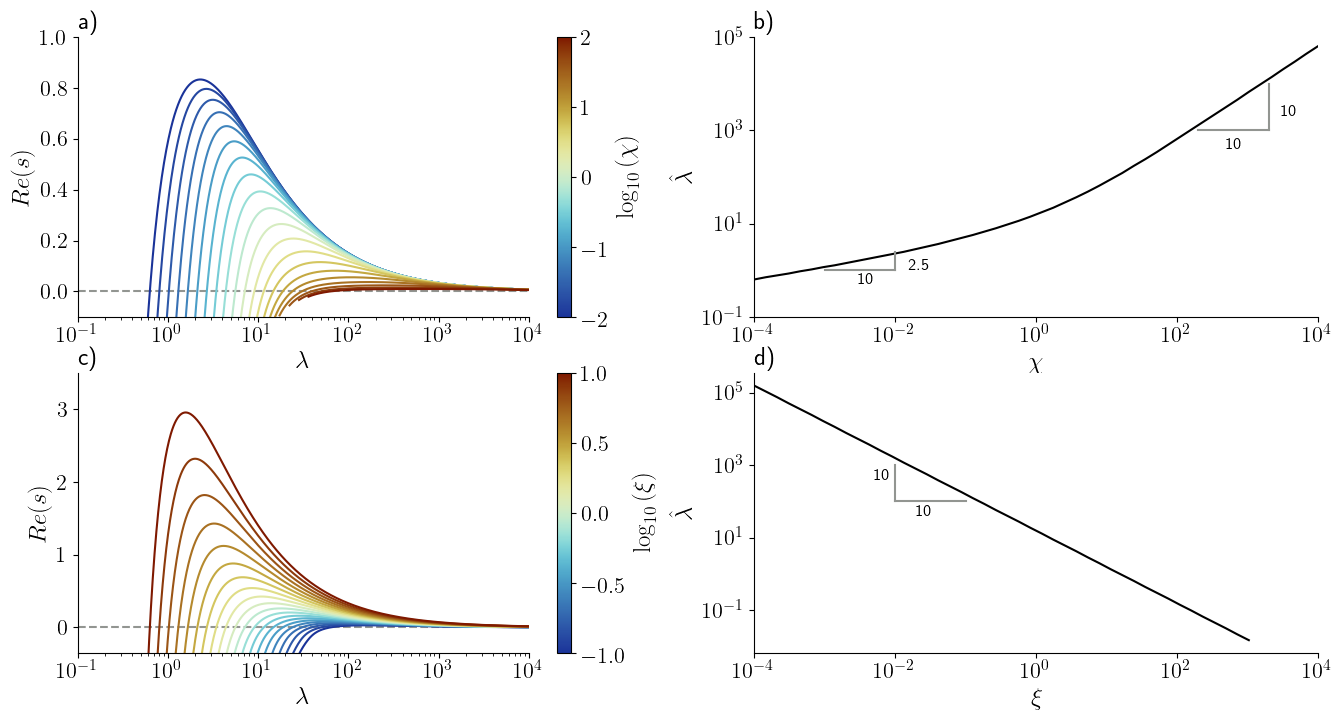}
  \caption{\label{fig:coupled}Impacts of diffusivity ratio \(\chi\) and time
  scale ratio \(\xi\) on the dominant wavelength of strain localization for a
  rate-dependent material subject to multiphysics softening only. Panels (a) and
  (c) show the results of the linear stability analysis for a range of values of
  \(\chi\) and \(\xi\) respectively. Panels (b) and (d) shows the dominant
  wavelength as a function of \(\chi\) and \(\xi\) respectively.}
\end{figure}

Figure~\ref{fig:coupled} illustrates the dependency of the dominant wavelength
\(\hat{\lambda}\) on the physical parameters and boundary conditions of the
problem for a material subject to multiphysics softening only. The impact of the
diffusivity ratio \(\chi\) (ratio of coupling diffusivity to momentum
diffusivity) is illustrated in panels (a) and (b) in Figure~\ref{fig:coupled}.
The dominant wavelength increases with the diffusivity ratio and a shift in the
slope can be observed when the coupling diffusivity is greater than the momentum
diffusivity (shift at \(\chi \sim 1\)). Below this critical value the
diffusivity ratio is controlled by momentum diffusivity and the perturbation
amplitude growths significantly faster (high values of \(Re\left(s\right)\))
than when the diffusivity ratio is dominated by the physical diffusivity of the
coupled process. Panels (c) and (d) in Figure~\ref{fig:coupled} illustrate the
impact of the time scale ratio \(\xi\) (ratio of the loading time scale to the
viscous time scale). The dominant wavelength is inversely proportional to the
time scale ratio (slope \(-1\) in \(\log\)-\(\log\) scale). Furthermore, for
high values of the time scale ratio \(\xi\) the perturbation grows significantly
faster. This effect can be interpreted in terms of the loading conditions: for
fast imposed strain rates, strain will localized slowly in a large extend
whereas for slow imposed strain rates, strain will localized fast into small
thicknesses. The findings presented here suggest that the introduction of
viscous terms are not sufficient to regularize the strain localization problem
(blue line in Figure~\ref{fig:uncoupled}). However, when considered together
with a softening process controlled by a coupled diffused variable, strain
localizes in finite thicknesses whose extents are physically controlled by the
diffusivity and time scale ratios. As deforming solid materials are usually
described with internal plastic hardening or softening, we investigate in the
following the different regimes obtained when internal and multiphysics
hardening/softening mechanisms interact or compete.

\subsection{Combined multiphysics coupling and strain hardening/softening}

In the following, we present the different responses of strain localization
depending on the values of the parameters for a material undergoing both
internal hardening/softening (\(h_{\gamma} \neq 0\)) and hardening/softening due
to multiphysics coupling (\(h_{\theta} \neq 0\)).

\begin{figure}
  \includegraphics[width=\textwidth]{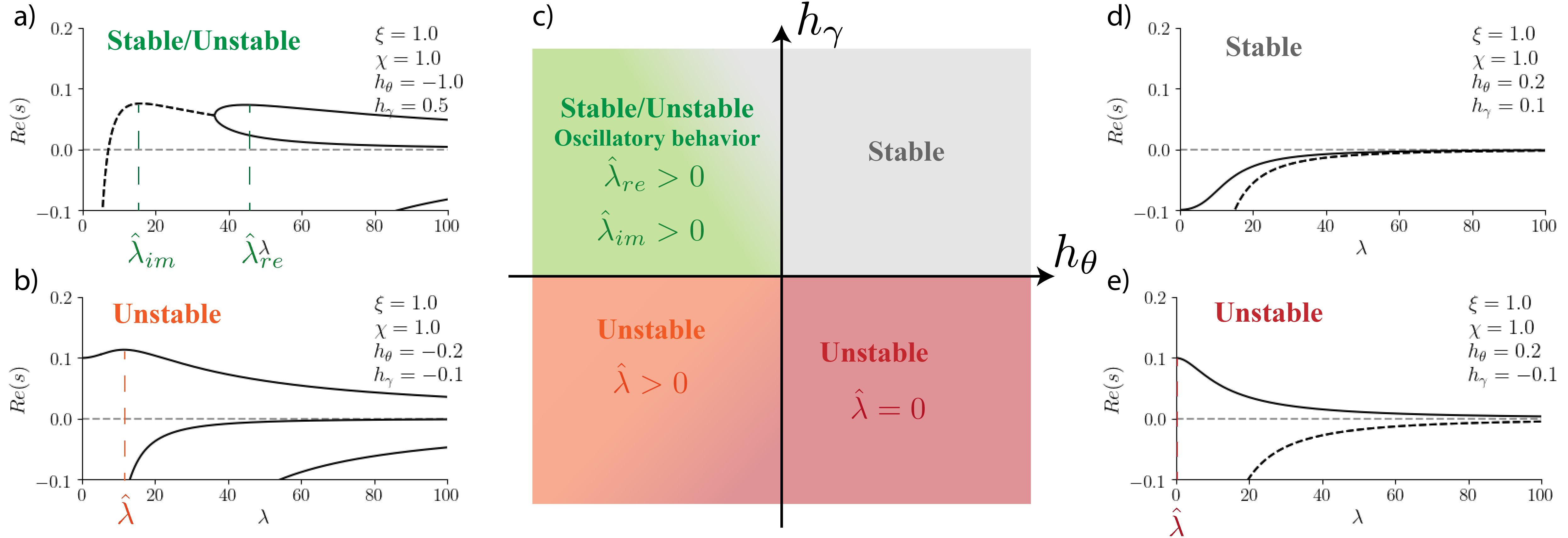}
  \caption{\label{fig:h_diagram}Overview of the results of the linear stability
  analysis. Panel (c) shows the stability conditions depending on the values of
  the two moduli \(h_{\theta}\) and \(h_{\gamma}\). Panels (a), (b), (d) and (e)
  show the real part of the rate of growth of the perturbation as a function of
  the perturbation wavelength for the representative regions described in panel
  (c). Solid lines indicate a pure real solution and dashed lines the real part
  of an imaginary solution. For each panel, we indicate the strain localization
  wavelength \(\hat{\lambda}\) defined as the wavelength for which the rate of
  growth of the perturbation is maximum. The strain localization problem is said
  to be regularized if \(0 < \hat{\lambda} < +\infty\).}
\end{figure}

Figure~\ref{fig:h_diagram} shows a map of the stability of the system depending
on the values of the two plastic moduli \(h_{\theta}\) and \(h_{\gamma}\)
together with representative linear stability analysis results for each case. If
both moduli are positive, the material only undergoes hardening and the material
behavior is always stable (gray region in panel (c) and panel (d)).  If the
material is subject to internal weakening and coupled hardening (\(h_{\gamma} <
0\) and \(h_{\theta} > 0\)), the behavior is unconditionally unstable and a
maximum bounded value of the real part of the rate of growth of the perturbation
is obtained for a wavelength equals to zero (red region in panel (c) and panel
(e)). Such system is characterized as being regularized in time because the
maximum value of the rate of growth of the perturbation is bounded but remains
unregularized in space as the dominant wavelength is zero. Similar results can
be obtained in the case when \(h_{\theta} = 0\) and \(h_{\gamma} < 0\) as
demonstrated by \citep{Stefanou2019} and from the previous section (see
Figure~\ref{fig:uncoupled}). When the material is subject to both internal
weakening and coupled weakening (\(h_{\gamma} < 0\) and \(h_{\theta} < 0\),
orange/red region in panel (c)), the system is unstable and can be regularized
both in time and space as depicted in panel (b). We cover in this contribution
the conditions necessary for regularizing the system in space, meaning that the
dominant wavelength \(\hat{\lambda}\) corresponding to the maximum of the rate
of growth of the perturbation is finite and non-zero (see panel (b)). The last
region, defined by competing internal hardening and coupled weakening
(\(h_{\theta} < 0\) and \(h_{\gamma} > 0\), green region in panel (c) and panel
(a)) give rise to a more complex behavior. Two local maxima of the real part of
the rate of growth of the perturbation can be identified: one corresponding to a
complex solution (dashed line in panel (a)) and a second one corresponding to a
pure real solution (solid line). The existence of a complex unstable solution
can be interpreted as an oscillatory instability. As the modulus \(h_{\gamma}\)
is increasing and the modulus \(h_{\theta}\) tending toward zero, the real part
of the rate of growth of the perturbation tends toward zero, hence exhibiting a
continuous transition toward stable conditions (gray area in
Figure~\ref{fig:h_diagram}-c). This regime is investigated in more details in
section~\ref{sec:pw}. In the following, we investigate the conditions for
regularizing strain localization for a system characterized by internal and
coupled weakening (panel (b)) and the underlying implications for the evolution
of strain localization.

\section{Strain localization regularization by standing waves (\(h_{\gamma} <
0\) and \(h_{\theta} < 0\))}
\label{sec:sw}

In this section, we cover the regime described by a overall weakening both in
the accumulated plastic strain (\(h_{\gamma} < 0\)) and in the coupled variable
(\(h_{\theta} < 0\)). As depicted in Figure~\ref{fig:h_diagram}, this regime is
unconditionally unstable but not necessarily regularized in a strain
localization sense. We therefore describe under which conditions, both in terms
of physical properties and boundary conditions, the strain localization
phenomenon is regularized both in time and space.

\subsection{Wavelength of strain localization}

\begin{figure*}
  \includegraphics[width=\textwidth]{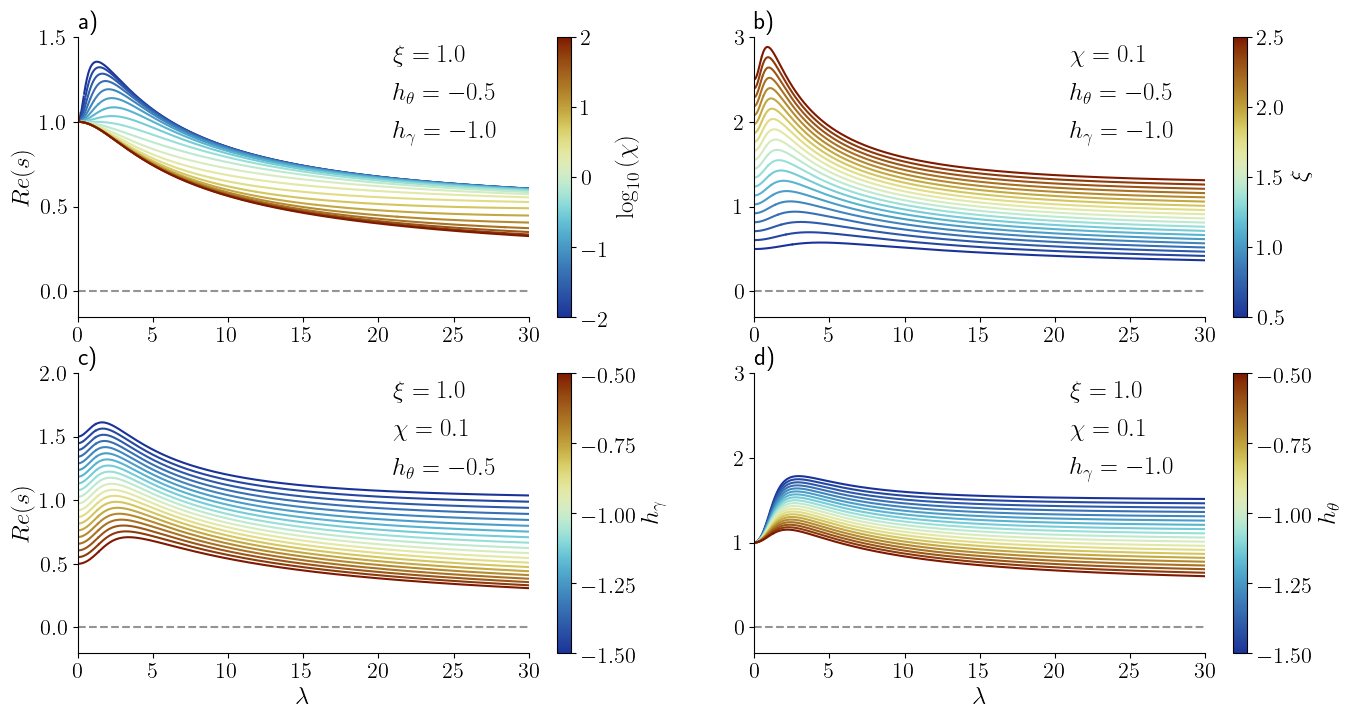}
  \caption{\label{fig:sw_params}Results of the linear stability analysis in
  response to a range of values for the four physical parameters: diffusivity
  ratio \(\chi\) (top left), time scale ratio \(\xi\) (top right), internal
  plastic modulus \(h_{\gamma}\) (bottom left) and plastic coupled modulus
  \(h_{\theta}\) (bottom right).}
\end{figure*}

Figure~\ref{fig:sw_params} shows the results of the linear stability analysis
for a regime characterized by softening both in plastic strain and in the
coupled diffused variable depending on a range of values for the different
physical parameters. We investigate the influence of the diffusivity ratio
\(\chi\), the time scale ratio \(\xi\) and the two dimensionless plastic moduli
\(h_{\gamma}\) and \(h_{\theta}\). For the given set of parameters, the results
suggest that it exists a critical value of the diffusivity ratio for which the
strain localization is regularized. For values of the diffusivity ratio
(\(\chi\)) higher than this critical value, the maximum value for the rate of
growth of the perturbation corresponds to a dimensionless wavelength of zero,
meaning that the strain localization problem is ill-posed (see red line in
Figure~\ref{fig:sw_params}-a). The regularization of the strain localization
problem by diffusive process occurs only for diffusivity ratios \(\chi\) below
this critical value (see blue line in Figure~\ref{fig:sw_params}-a). In
addition, decreasing the diffusivity ratio decreases the dominant wavelength of
strain localization. The time scale ratio (\(\xi\)) does not have any impact on
the regularization of the strain localization problem but only influence the
absolute value of the rate of growth of the perturbation and its dominant
wavelength. A low value of the time scale ratio \(\xi\) corresponds to high
viscous strain rates which leads to a fast growing perturbation. The effects of
the plastic modulus \(h_{\gamma}\) are similar to those of the effective strain
rate ratio \(\xi\). Increasing the plastic moduli however increases the value of
the regularized wavelength. Finally the coupled plastic modulus \(h_{\theta}\)
influences the long wavelengths but has limited influences to short ones.
Increasing the coupled plastic modulus leads to a decrease in the strain
localization wavelength.

%

\begin{figure*}
  \includegraphics[width=\textwidth]{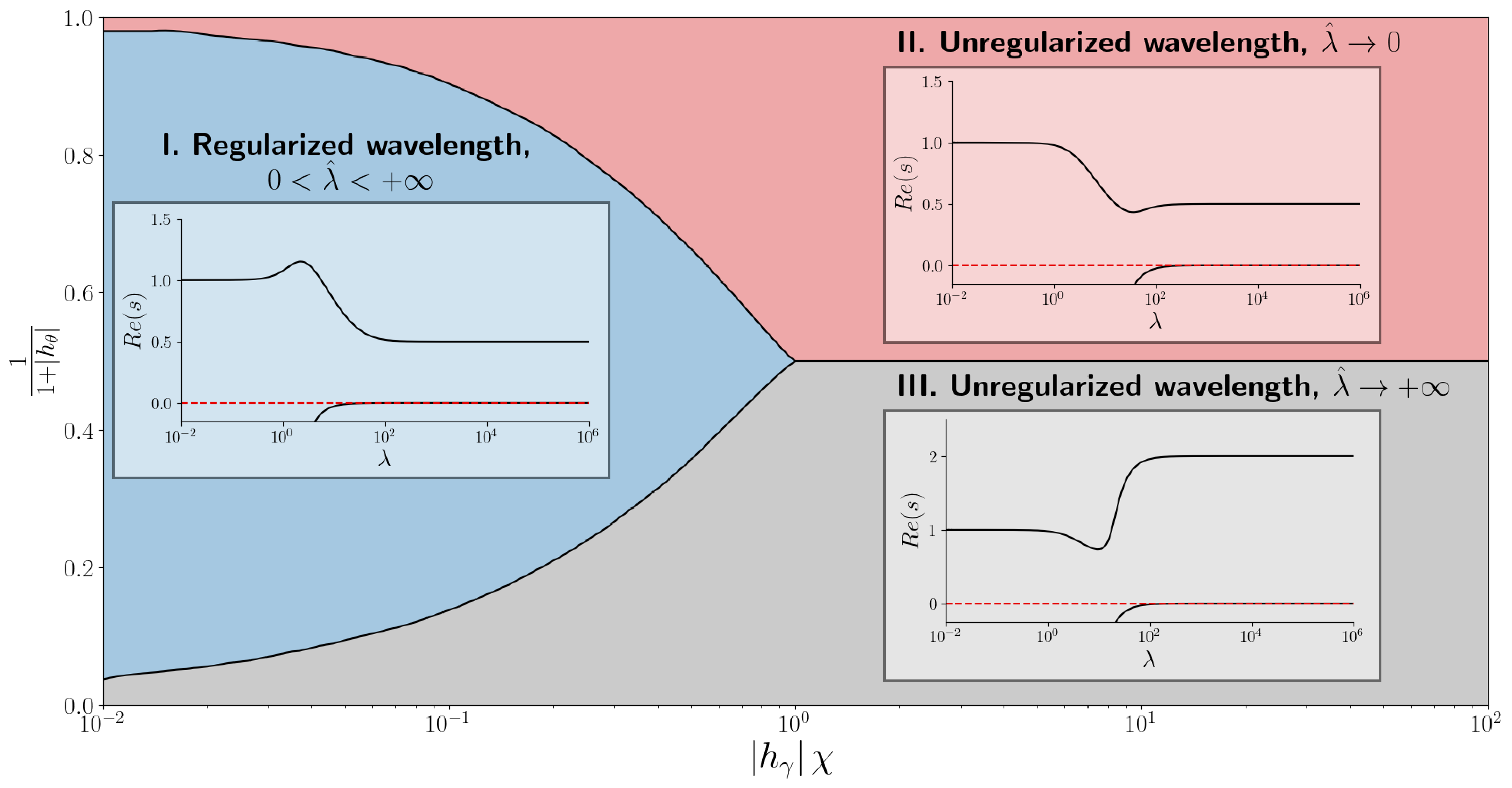}
  \caption{\label{fig:sw_regimes}Phase diagram for the strain localization
  problem. The different regimes are displayed as a function of the effective
  diffusivity ratio \(\left|h_{\gamma}\right| \chi\) and the effective coupled
  plastic modulus \(\frac{1}{1 + \left|h_{\theta}\right|}\). The blue regime
  refers to a regularized strain localization problem whereas the red and gray
  regimes refer to an unregularized problem for which the dominant wavelength
  either tends toward zero or infinity respectively.}
\end{figure*}

Figure~\ref{fig:sw_regimes} illustrates the different regimes obtained by
appropriately scaling all the parameters (\(h_{\theta},h_{\gamma},\chi,\xi\))
involved in the strain localization problem. We introduced the effective
diffusivity ratio \(\left|h_{\gamma}\right|\chi\) defined as the product of the
diffusivity ratio and the absolute value of the internal weakening modulus and
the effective coupled plastic modulus \( \frac{1}{1 + \left|
h_{\theta}\right|}\) as a measure of the coupled weakening process. The
regularization via diffusive process of the strain localization problem is only
effective when the effective diffusivity ratio is smaller than one (see regime I
in blue). For an overdiffused system, or for large internal weakening mechanism,
the system is in general unregularized. In that case, for mild coupled weakening
mechanism (low value of \(h_{\theta}\)), the strain localization wavelength
tends toward zero (regime II in red) and for large coupled weakening mechanism
(high value of \(h_{\theta}\)), it tends toward infinity (regime III in gray). A
general condition for the regularization of the strain localization problem is
therefore a low diffusivity of the internal diffusive process together with
competing weakening mechanisms. In general, it can be concluded that internal
weakening tends to unregularize the strain localization problem whereas coupled
weakening favors an internal length scale. It is also worth noticing that our
results suggest that the regime described in Figure~\ref{fig:sw_regimes} are
independent from the time scale ratio \(\xi\) and thus from the loading strain
rate. However, this quantity impacts the obtained length scale of the strain
localization as described in Figure~\ref{fig:coupled}.


\subsection{Evolution of strain localization}

To understand the implications of the existence of a regularized wavelength for
strain localization, we performed dynamic simulations using the finite-element
method. We implemented the system of equations described in
Equation~\ref{eq:final_linearized} relying on the MOOSE environment
(Multiphysics Object-Oriented Simulation Environment) \citep{Permann2020}. MOOSE
is an open source environment for multiphysics Finite Element applications
developed by the Idaho National Laboratories (INL,
\url{https://mooseframework.inl.gov/}). It provides a flexible, hybrid parallel
platform to solve for multiphysics and multicomponent problems in an implicit
manner.

For this analysis, we consider a one-dimensional bar of dimensionless length
\(100\) subject to background strain rate \(\dot{\gamma}_{0}\). As we are
solving for the perturbations of the state variables, the displacement
perturbation variable \(\bar{u}^{\star}\) is initially homogeneous and equals to
zero. The coupled variable perturbation \(\theta^{\star}\) is initialized by
considering a perturbation at the center of the domain given by the following
equation:

\begin{equation}
  \theta^{\star}\left(\bar{x}, \bar{t}=0\right) =
  \begin{cases}
    \frac{1}{2} \theta^{\star}_{0}\left(1 + \cos\left(\frac{2 \pi
    \bar{x}}{\lambda_{0}}\right)\right), & \text{ if } -\frac{\lambda_{0}}{2} <
    \bar{x} < \frac{\lambda_{0}}{2},\\
    0 & \text{otherwise.}
  \end{cases}
\label{eq:perturbation_fem}
\end{equation}

where \(\theta^{\star}_{0}\) and \(\lambda_{0}\) are the initial perturbation
amplitude and wavelength respectively. The solution given by the finite-element
analysis allows to describe the distribution and evolution with time of the
displacement and coupled variable perturbation fields. We consider here the case
where the physical parameters give a regularized wavelength as depicted for
regime I in Figure~\ref{fig:sw_regimes}.

\begin{figure}
  \centering
  \includegraphics[width=0.75\columnwidth]{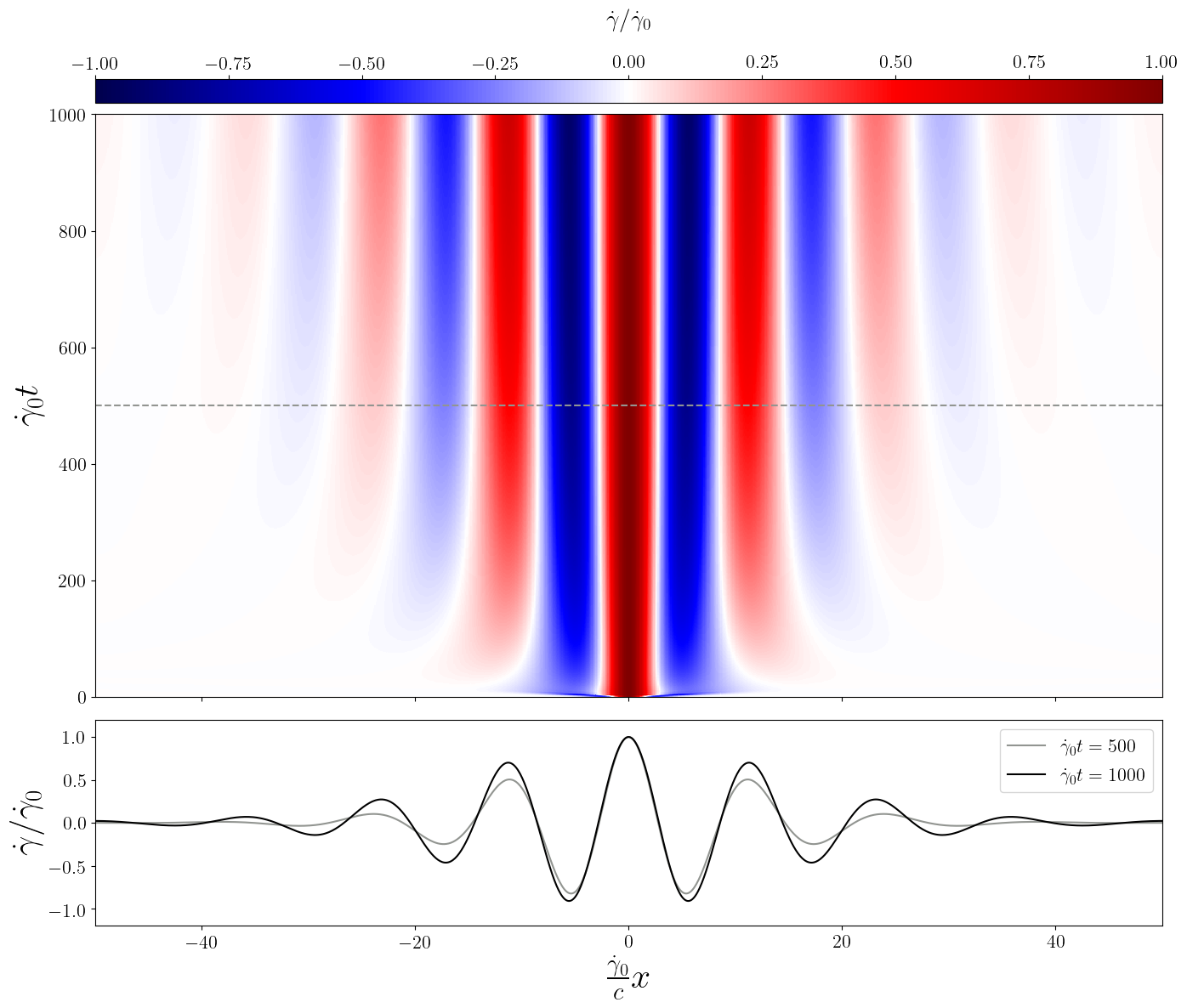}
  \caption{\label{fig:fem_regularized}Evolution and distribution of the
  perturbation evaluated by finite-element analysis. The wavelength of the
  initial coupled variable perturbation is taken as \(\lambda_{0} = 5.0\). Top
  panel shows the distribution and evolution of the total strain rate. Bottom
  panel shows the distribution along the bar of the total strain rate at two
  different times. As the perturbations grows exponentially with time,
  normalized values are plotted.}
\end{figure}

\begin{figure}
  \centering
  \includegraphics[width=0.75\columnwidth]{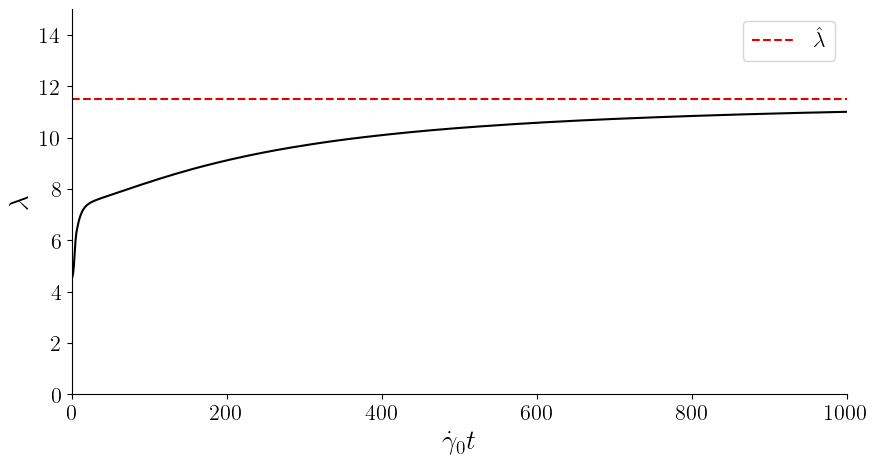}
  \caption{\label{fig:fem_wl_regularized}Evolution with time of the perturbation
  wavelength obtained from the finite-element simulations. The red dotted line
  indicates the dominant wavelength obtained from the linear stability
  analysis.}
\end{figure}

Figure~\ref{fig:fem_regularized} shows the distribution and evolution in time of
the strain rate perturbation for a regularized system. The initial perturbation
in the coupled variable perturbation field drives the system to strain
localization. The wavelength of the strain localization gradually evolves toward
the value obtained from the linear stability analysis. As the initial wavelength
of the perturbation is smaller than the internal wavelength of the system, the
size of the strain localization increases progressively upon reaching the
internal wavelength as depicted in Figure~\ref{fig:fem_wl_regularized} showing
the evolution of the strain localization wavelength with time. As the
perturbation evolves over time, it generates regular patterns governed by the
strain localization wavelength.

%
%

\section{Strain localization regularization by standing and propagating waves
(\(h_{\gamma} > 0\) and \(h_{\theta} < 0\))}
\label{sec:pw}

In this section we investigate the case where deformation is characterized by
competing hardening and softening. In particular, we consider the case where the
system undergoes hardening in the plastic strain (\(h_{\gamma} > 0\)) and
softening in the coupled variable (\(h_{\theta} < 0\)). As depicted in
Figure~\ref{fig:h_diagram} panel (a), this regime is characterized by two
dominant wavelengths, the first one being attributed to standing waves and the
second one to propagating waves. Here we investigate the conditions in terms of
physical parameters and boundary conditions which allow these two sets of waves
to interplay.

\subsection{Standing and propagating waves}

\begin{figure*}
  \includegraphics[width=\textwidth]{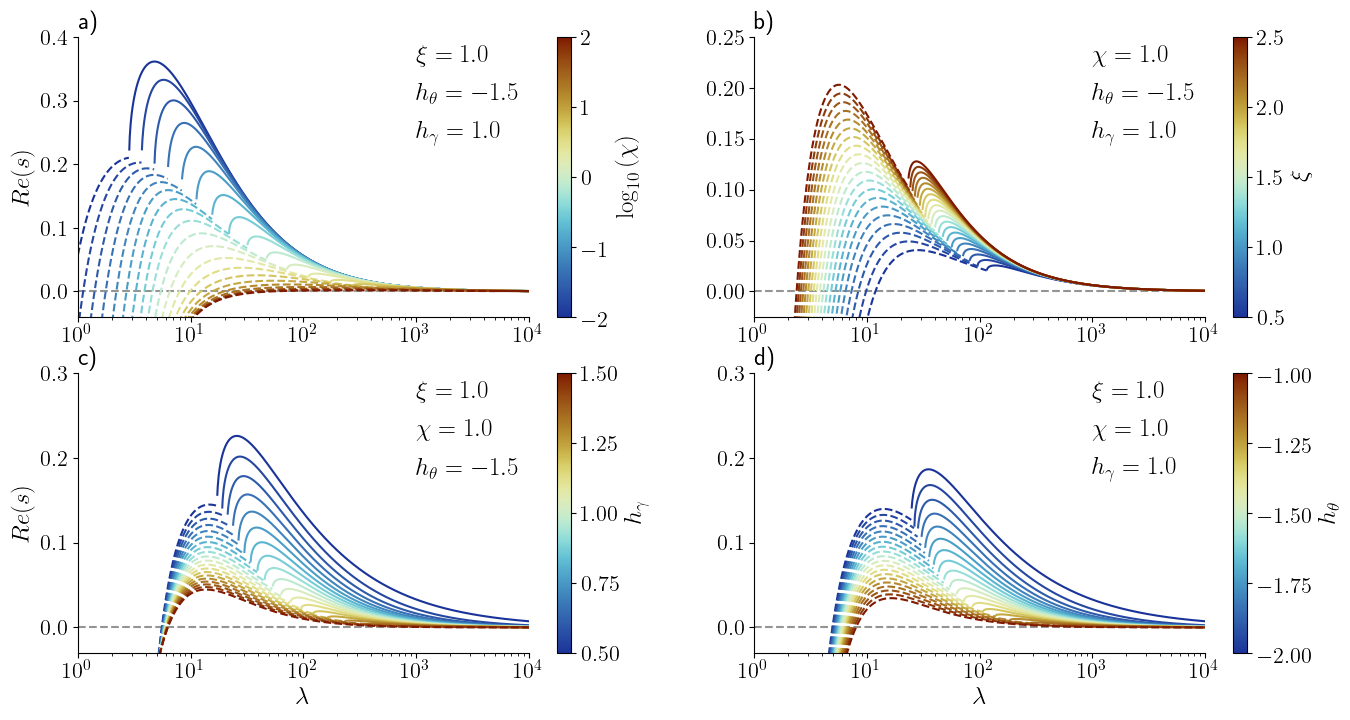}
  \caption{\label{fig:ps_params}Results of the linear stability analysis in
  response to a range of values for the four physical parameters: diffusivity
  ratio \(\chi\) (top left), time scale ratio \(\xi\) (top right), internal
  plastic modulus \(h_{\gamma}\) (bottom left) and plastic coupled modulus
  \(h_{\theta}\) (bottom right).}
\end{figure*}

Figure~\ref{fig:ps_params} illustrates the results of the linear stability
analyses for a range of values of the four parameters (\(\chi, \xi, h_{\gamma},
h_{\theta}\)). The perturbations evolve toward two different wavelengths upon
strain localization as presented in Figure~\ref{fig:ps_params}. The first one
corresponds to an oscillatory regime (real part of a complex solution, dashed
lines in Figure~\ref{fig:ps_params}) which can be characterized by propagating
waves. The second one (real solution, solid lines in Figure~\ref{fig:ps_params})
corresponds to standing waves and is often referred as the spacing or the
thickness of localized bands (see previous section and
Figure~\ref{fig:fem_regularized}). The coexistence of these two sets of waves
highly depends on the parameter values which also dictate which regime dominates
the strain localization phenomenon. An interesting aspect of these findings is
that the diffusivity ratio \(\chi\) seems to play a major role in switching
between dominating standing waves (for low values of \(\chi\)) to dominating
propagating waves (for high values of \(\chi\)) which captures the competing
effect of physical to momentum diffusivities. The coupled diffusive process
therefore controls how strain localization evolves in a deforming solid and the
nature of its macroscopic manifestations. While the effect of the time scale
ratio \(\xi\) does not seem to influence the existence of standing and
propagating waves, the diffusivity ratio \(\chi\) and the two plastic moduli
\(h_{\gamma}\) and \(h_{\theta}\) play a major role in controlling which waves
(standing or propagating) are dominant. The competing effect of internal
hardening and coupled softening is also relevant to the transition between the
two regimes, which is well captured by the effective coupled plastic moduli
\(\frac{1}{1 + \left|h_{\theta}\right|}\) bounded between zero and one and the
effective diffusivity ratio \(\left|h_{\gamma}\right| \chi\) as introduced in
section~\ref{sec:sw}.

\begin{figure*}
  \includegraphics[width=\textwidth]{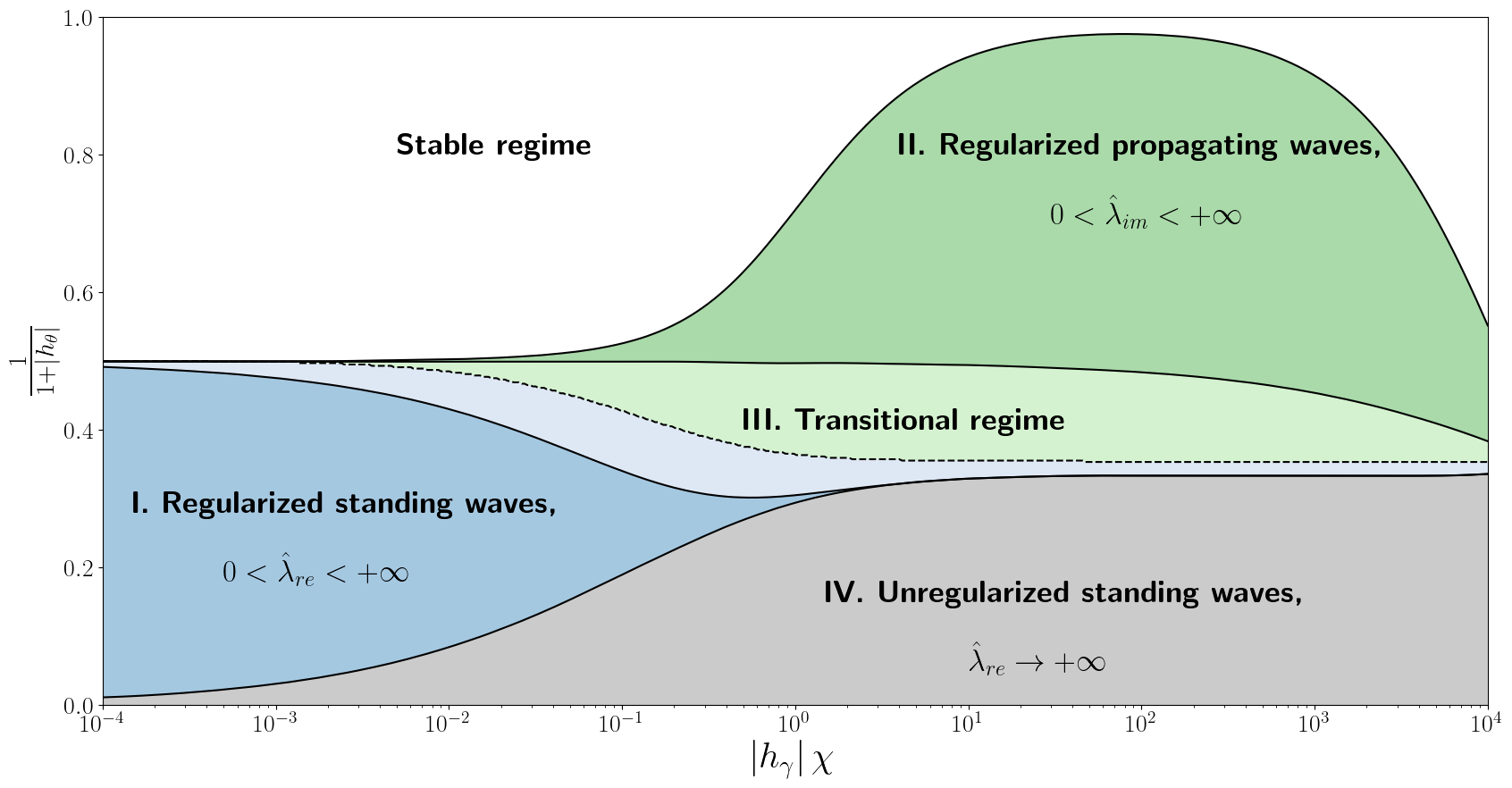}%
  \caption{\label{fig:pw_regimes}Phase diagram for the different strain
  localization regimes as a function of the effective diffusivity ratio
  \(\left|h_{\gamma}\right| \chi\) and the effective coupled plastic modulus
  \(\frac{1}{1 + \left|h_{\theta}\right|}\). Regime I in blue corresponds to
  regularized standing waves, regime II in green to regularized propagating
  waves. Regime III is a transitional regime (light blue and light green) where
  the two wavelengths (standing and propagating waves) coexist. Light green
  region indicates that the propagating waves grow faster than the standing
  waves and the light blue region the opposite. Regime IV in gray indicates a
  non-regularized standing waves regime, similar to the gray regime in
  Figure~\ref{fig:sw_regimes}. The white region indicates a stable behavior
  where the perturbations decrease over time.}
\end{figure*}

Figure~\ref{fig:pw_regimes} illustrates the different regimes we identified in
the \(\left|h_{\gamma}\right| \chi\), \(\frac{1}{1 + \left|h_{\theta}\right|}\)
space for a system undergoing internal hardening and coupled weakening. Standing
waves are obtained when the coupled weakening mechanism dominates over the
internal hardening mechanism and for low diffusivity ratio (high \(h_{\theta}\)
and low \(\left|h_{\gamma}\right| \chi\)). As the internal hardening or
diffusivity increases, propagating waves can emerge, coexisting with the
standing waves (transitional regime III as depicted in
Figure~\ref{fig:pw_regimes}). For low coupled weakening mechanism, standing
waves can fade away and only propagating waves remain (regime II). For low
coupled weakening mechanism at low effective diffusivity ratio, a stable regime
is obtained where the perturbation fades away with time to go back to the
steady-state solution. Similar to the findings presented in
section~\ref{sec:sw}, an unregularized standing waves regime is obtained for
high coupled weakening mechanism leading to infinite wavelengths (regime IV in
Figure~\ref{fig:pw_regimes}).

\subsection{Evolution of strain localization with interacting standing and
propagating waves}

\begin{figure}
  \centering
  \includegraphics[width=0.75\columnwidth]{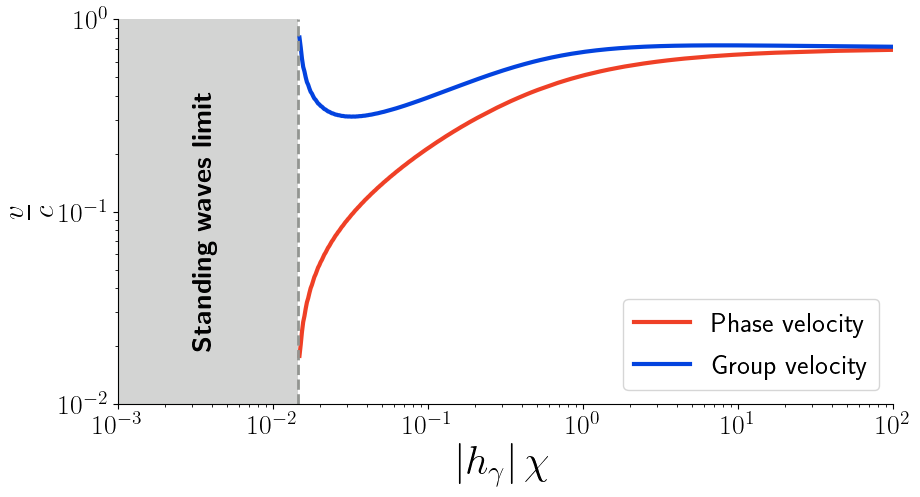}%
  \caption{\label{fig:velocities}Evolution of the ratio of phase and group
  velocities to the elastic wave velocity as a function of the effective
  diffusivity ratio \(\left|h_{\gamma}\right| \chi\) obtained with \(h^{\theta}
  = -1.5\). Below a critical value of the effective diffusivity ratio, the
  standing waves regime is reached and the phase velocity tends toward zero and
  the group velocity toward infinity.}
\end{figure}

Figure~\ref{fig:velocities} shows the evolution of the phase and group
velocities (normalized by the elastic wave velocity) as a function of the
effective diffusivity ratio \(\left|h_{\gamma}\right| \chi\) as predicted by the
linear stability analysis. Below a critical value of diffusivity, propagating
waves vanish and the standing waves regime is reached. Approaching this regime,
the group velocity tends toward infinity and the phase velocity to zero. The
phase velocity is always lower than the group velocity but both converge to a
constant value for high values of \(\left|h_{\gamma}\right| \chi\). The strain
localization waves velocity remains always smaller than the elastic wave
velocity. These findings suggest that even in the case of propagating strain
localization activated by diffusive process, the strain localization waves
propagate slower than the elastic waves. In this case, the elastic wave velocity
acts as a limit for strain localization propagation.

\begin{figure*}
  \includegraphics[width=\textwidth]{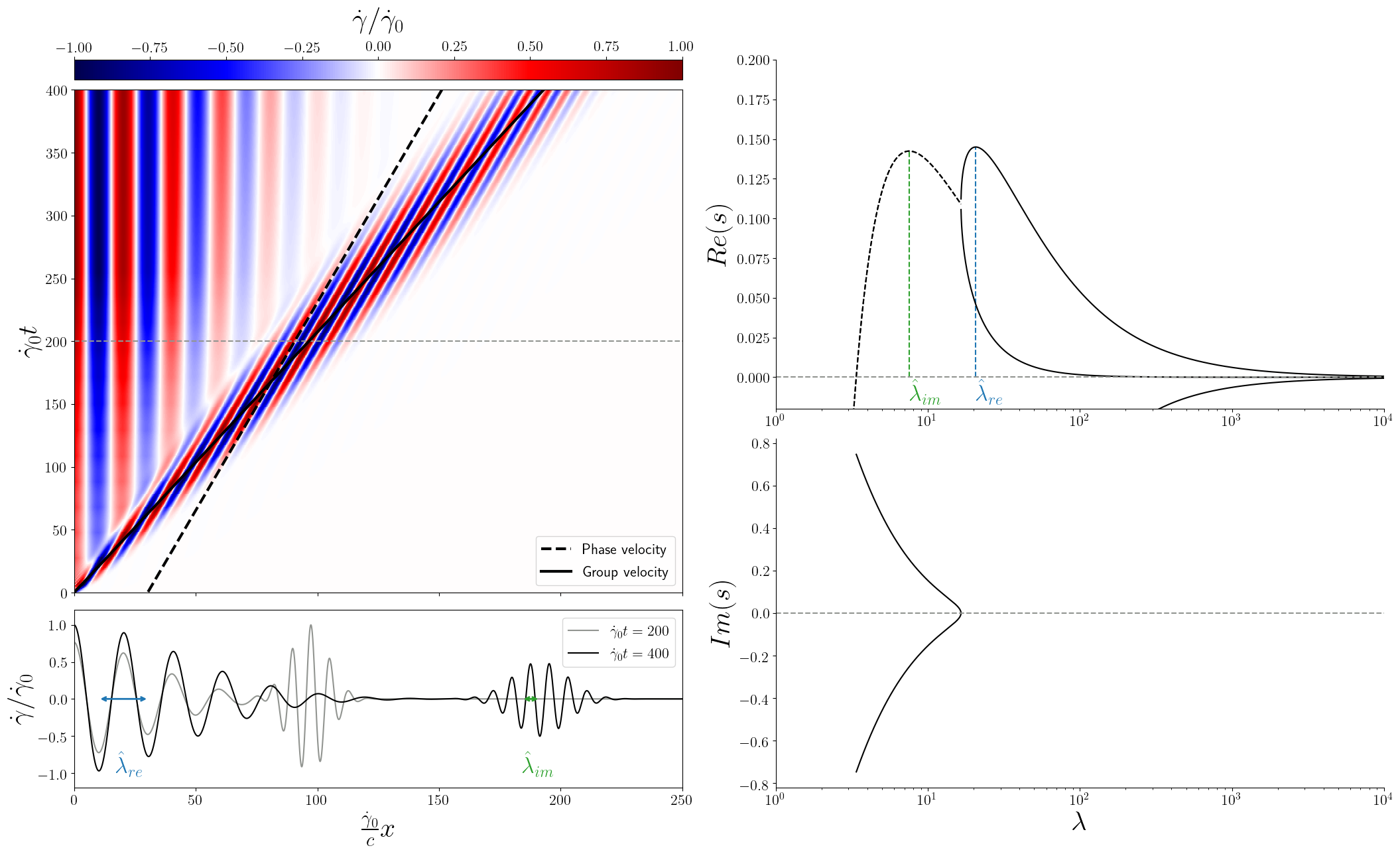}%
  \caption{\label{fig:fem}Left panels show the evolution and distribution of the
  normalized total strain rate for a dynamic simulation. The predicted phase and
  group velocities from the linear stability analysis are indicated as well as
  the two wavelengths corresponding to standing and propagating waves. Right
  panels show the results from the linear stability analysis for the set of
  parameters chosen in the dynamic simulation.}
\end{figure*}

We performed some dynamic simulations for this regime using the
finite-element-method based MOOSE framework \cite{Permann2020} to analyze the
evolution of the strain localization. We cover the case of the standing waves
regime in some details in section~\ref{sec:sw} and focus here on the
transitional regime when both standing and propagating waves coexist. We
consider a similar setting and the same initial perturbation in the coupled
variable \(\theta^{\star}\) (Equation~\ref{eq:perturbation_fem}) as presented in
section~\ref{sec:sw} for a dimensionless bar of length \(500\) centered in zero
(we report here only the results of the positive half as the results are
symmetric).

Figure~\ref{fig:fem} shows the distribution of the normalized total strain rate
over time together with the results of the linear stability analysis for the
system considered. The initial imposed perturbation generate a propagating wave
characterized by a group velocity larger than the phase velocity as predicted by
the linear stability analysis. Behind this propagating front, standing waves
develop which self organize in a regular way. The two coexisting wavelengths
predicted by the linear stability analysis match with the thickness of the
standing and propagating waves. The macroscopic manifestation of such a
transitional regime can be seen as an erratic distribution and propagation of
deformation due to the discrepancy between phase and group velocities, followed
by the formation of organized deformation patterns. The group and phase
velocities observed in the simulations also match the predictions obtained from
the linear stability analysis.

\section{Discussion}


In this study, we demonstrated under which conditions the introduction of a
viscous term or the use of a viscoplastic model can successfully lead to the
regularization of strain localization phenomena. The combination of
viscoplasticity together with the multiphysics coupling to a diffusive process
allows to carry the diffusive length scale to the deformation constitutive laws
and therefore obtain a physical length scale for strain localization. The
concept of viscous regularization has been widely studied in the modeling
community. Several contributions investigated numerically the effect of
viscoplasticity on the localization of the displacement or velocity fields with
some success in terms of regularization of the ill-posed problem
\citep{Needleman1988,Loret1990,Prevost1990,Duretz2014,Duretz2019,Jacquey2020}
but always missed an analytical or mathematical proof that the length scale
observed in the numerical results uprose from the constitutive laws. The results
presented in this contribution argue that viscous terms alone do not regularize
the strain localization in space but only in time. The artificial regularization
observed in numerical models therefore is the results of a regularization in
time from the viscous terms which allows to carry the length scale (usually
quite large) introduced by the initial perturbation of the strain field. The
length scale obtained in such models does not refer to a physical length scale
but rather depends on the initial conditions, and thus on the size of the
perturbation introduced. We report in this contribution that (i) viscous terms,
(ii) inertia effects and (iii) multiphysics coupling are necessary to obtain a
physical length scale for strain localization. Weakening mechanisms triggered by
a coupled diffusive variable therefore tend to regularize the strain
localization problem whereas internal weakening mechanisms promote unregularized
internal length scale of zero size. The successful regularization of the strain
localization problem, when the perturbation wavelength is positive, finite and
determined by physical parameters depends on the competition between these two
weakening mechanisms. While the loading rate conditions influence the size of
the dominant wavelength and thus the internal length scale, they do not have any
impact on the regularization conditions.

An essential aspect in the formulation we propose in this contribution is the
coupling between the deformation mechanism and a diffusive process. This
coupling allows to integrate the diffusive length-scale into the deformation
localization phenomenon. As demonstrated in the present study, the dominant
length-scale upon localization highly depends on the effective diffusivity of
the material. This coupling is therefore essential to regularize the ill-posed
problem of deformation localization. Such coupling is in general common for the
rheological description of solid materials. Several studies considered
thermo-mechanical \citep{Veveakis2014b,Paesold2016} or chemo-mechanical
\citep{Brantut2011,Buscarnera2012,Stefanou2014,Sulem2016} coupling as playing an
important role for the regularization of strain localization. For geomaterials,
generally described as porous media, the diffusion of pore pressure and
saturation was also considered in the literature
\citep{Zhang2000,Schrefler2006,Veveakis2015b,Alevizos2017,Oka2019}. The
introduction of coupled diffusive process is also at the base of non-local
theories such as gradient damage or breakage mechanics \citep{Nguyen2010}. This
study demonstrates how the coupling to a single diffusive process generates
long-lasting localization patterns. Several studies considered the deformation
of solid materials as the result of complex multiphysics involving several
couplings to diffusive processes \citep{Segall2006,Platt2014}. Such approaches,
described as THMC (thermal-hydro-mechanical-chemical) coupling imply the
existence of several length scales associated to the different diffusive
processes considered \citep{Veveakis2015}. The overall behavior of multiphysics
solid materials can generally vary over the scale considered as each diffusive
process is dominant at a specific length and time-scale. In this regard, the
results presented in this contribution considering the coupling to a single
diffusive process can be used to describe the multiphysics of deforming solid
materials at a specific scale at which the chosen diffusive process is dominant.
Extending the current approach to multiple diffusive processes, thus considering
the multiscale aspect of the deformation of solid materials could be the topic
of future studies.

We covered in this contribution two specific regimes which are characterized by
the formation of regular patterns due to standing waves and by the formation of
erratic patterns due to oscillatory solutions and the interplay of propagating
and standing waves. As we solved for the dynamic evolution of a linearized
system -- we did not consider a particular form of the shear stress function but
rather considered its derivatives evaluated at the steady-state solution -- the
patterns obtained are periodic in space and we could identify well-defined
regimes. Considering a particular non-linear shear stress function would
eventually not lead to similar periodic patterns as the derivative of the shear
stress function are not constant over time in a general sense. This suggests
that the deformation of a material characterized by a non-linear shear stress
(which is generally the case) would not be limited to one of the particular
regimes we described here but rather could evolve in time and therefore results
in transient dominant wavelengths and a potential switch from erratic to regular
pattern over time. In particular, most materials undergo hardening in the
internal plastic strain after yielding and gradually transition to a softening
regime upon continuous loading. Following the results presented here, this would
imply that the dissipative propagating waves may emerge shortly after yielding
but would vanish over time and let standing waves dominating the deformation at
a larger time scale. The results we presented in this contribution could
therefore have implications for the deformation of granular materials for which
such waves have been observed in the laboratory \citep{Guillard2015}, for the
initiation and propagation of earthquake rupture characterized by velocities
lower than the elastic wave velocity \citep{Kanamori1994}, slow slip events
\citep{Dragert2001,Obara2004} or for the driving mechanisms responsible for the
Portevin-Le Chatelier effect in steel and aluminum alloys which is characterized
by the progressive formation of regular patterns \citep{Yilmaz2016}.

\section{Conclusion}

In this contribution, we investigate the behavior of a one-dimentional
rate-dependent plastic formulation coupled to a diffusive process for deforming
solid materials. We demonstrate the conditions in terms of physical parameters
to regularize both in space and time strain localization using a linear
stability analysis. We demonstrated that inertia effects together with the
rate-dependency of the plasticity model adopted give minimum values for the
applied strain rate and physical diffusivity values to effectively regularize
the strain localization problem. Using finite-element analysis, we illustrated
how the strain localization wavelength evolves over time in response to a given
perturbation. The results described in this contribution are relevant to the
multiscale description of localization phenomena in a large variety of solids
exhibiting shear or compaction bands. As we considered a generic coupled
diffusive process, the results presented in this contribution are extendable to
any multiphysical system such as thermo-mechanical, chemo-mechanical or
hydro-mechanical coupling but also to non-local theories such as gradient damage
or breakage mechanics. We also investigated using the same set of governing
equations, the existence of an oscillatory regime with two wavelengths with
implications for the propagation of strain localization in the form of
dissipative waves  and the formation of erratic patterns.


\section*{Acknowledgement}
A. B. Jacquey acknowledges support by the Helmholtz Association in the frame of
the Doctoral Award 2018 in the field of energy. M. Veveakis acknowledges support
by the DE-NE0008746-DoE, United States project.

\bibliography{references}

\end{document}